\documentclass[prd, superscriptaddress, preprintnumbers, showpacs]{revtex4}
\usepackage{epsfig}
\usepackage{graphicx}
\usepackage{amsmath}
\usepackage{bm}
\usepackage{amssymb}
\usepackage{color}

\newif\ifold             \oldtrue            
\def\ba{\begin{eqnarray}}
\def\ea{\end{eqnarray}}

\def\bp{\mathbf{p}}
\def\bk{\mathbf{k}}

\newcommand{\be}{\begin{equation}}
\newcommand{\ee}{\end{equation}}
\def\ba{\begin{eqnarray}}
\def\ea{\end{eqnarray}}
\def\bea{\begin{eqnarray}}
\def\eea{\end{eqnarray}}

\def\bp{\mathbf{p}}
\def\bk{\mathbf{k}}

\newcommand{\refer}[1]{(\ref{#1})}

\newcommand{\tr}{{\mbox{tr}}}

\begin{document}


\title{Gap generation  in ABC-stacked multilayer graphene: screening vs band flattening}
\date{\today}

\author{Junji Jia}
\email{junjijia@whu.edu.cn}
\affiliation{School of Physics and Technology, Wuhan University, 430072, Wuhan, China}

\author{E.V. Gorbar}
\email{gorbar@bitp.kiev.ua}
\affiliation{Department of Physics, Taras Shevchenko National Kiev University, 03022, Kiev, Ukraine}
\affiliation{Bogolyubov Institute for Theoretical Physics, 03680, Kiev, Ukraine}

\author{V.P. Gusynin}
\email{vgusynin@bitp.kiev.ua}
\affiliation{Bogolyubov Institute for Theoretical Physics, 03680, Kiev, Ukraine}

\begin{abstract}
We calculate the dynamical polarization function and solve a self-consistent gap equation in
the random phase approximation  in undoped ABC-stacked $n$-layer graphene. We find that the
gap is maximal in  trilayer graphene and decreases monotonously for $n \ge 4$ because the
effects of screening in the gap equation win over those connected with the flattening of
electron bands as $n$ increases.
\end{abstract}


\maketitle
\section{Introduction}
\label{1}

Multilayer graphene is a very special material in 2D electron systems. Its physical properties
strongly depend on the stacking order. Among many possibilities of stacking order at fixed number
of layers $n$, ABC-stacked or rhombohedral graphene has the most flat electron spectrum at low
energy. According to Refs.\cite{Guinea,Min}, this material belongs to a new class of two-dimensional electron systems (2DES) which is known as chiral 2DES \cite{Barlas} due to the chiral properties
of its low energy electron Hamiltonian.

Neglecting the trigonal warping effects (for a discussion of these effects and the electron spectrum
in multilayer graphene, see Refs.\cite{Koshino,Sahu}) the electron energy in chiral multilayer graphene
with $n$ layers at low energy is given by $\varepsilon(\mathbf{p}) \sim |\mathbf{p}|^n$. For example,
the low energy electron spectrum in bilayer graphene with Bernal stacking is characterized by two
parabolic bands touching at two points. Note that the above mentioned formula for $\varepsilon(\mathbf{p})$ is valid for single layer graphene too, where, as it is well known, $\varepsilon(\mathbf{p}) \sim |\mathbf{p}|$. ABC-stacked trilayer graphene has the low energy electron spectrum with cubic bands touching at two points. However, there exist different symmetry breaking terms relevant at low energy in trilayer graphene which essentially modify the cubic spectrum at energies $\lesssim 10$  meV.

A sizable gap can be opened in ABC-stacked trilayer graphene subjected to a perpendicular electric
field \cite{Avetisyan,Bao,Heinz,Sahu,Clapp} (gap opening and gate-tunable band structure in few-layer graphene are discussed in Refs.\cite{McCann,Russo}). Broken symmetry states in ABC-stacked trilayer
graphene were studied in a self-consistent Hartree--Fock approximation \cite{Jung}.
Ferromagnetism was considered in Ref. \cite{Olsen}, where the screening effects were taken into account within a simplified model. The effects of topology and electron-electron interactions on the phase
diagram of ABC-stacked trilayer graphene were investigated in Ref. \cite{Vafek}. The generation of
stacking order dependent gap in trilayer graphene was studied within a Hubbard model \cite{Yao}.
The integer quantum Hall effect in a gated trilayer graphene in high magnetic fields was experimentally observed \cite{Kumar} and the quantum Hall states in ABC-stacked trilayer and few-layer graphene
were considered in Refs.\cite{Niu,Cote,Toke}.

Clearly, as $n$ increases, the low energy electron spectrum $\varepsilon(\mathbf{p}) \sim |\mathbf{p}|^n$
becomes more flat in chiral multilayer graphene. The interaction parameter $r_s$, defined as the ratio of the average of inter-electron Coulomb interaction energy to the Fermi energy, scales like
$r_s\sim n^{(1-n)/2}_{\rm el}$ \cite{Sarma}, where $n_{\rm el}$ is the electron charge density. Obviously, the
electron-electron interactions become more essential at low electron density as the number of layers $n$ increases in ABC-stacked multilayer graphene. This suggests that the gap generation in chiral multilayer graphene should be enhanced \cite{Polini,Sun} as the number of layers $n$ becomes larger.

This conclusion seems to be experimentally confirmed for small $n$. Meanwhile no gap is observed in monolayer graphene at the neutrality point in the absence of external electromagnetic fields, a gap
$2$ meV is reported in bilayer graphene \cite{Martin,Weitz,Freitag,Velasco}. Still larger gap up to
$6$ meV is found in ABC-stacked trilayer graphene \cite{Bao}. The most recent experiments \cite{Lee,LeRoy} demonstrate the presence of gaps of almost room temperature magnitude $\sim 25$ meV in high mobility ABC-stacked trilayer graphene. Theoretically, the functional renormalization group study \cite{Honerkamp} found that the energy scale for the instability of the system with respect to
the gap generation comes out in ABC-stacked trilayer graphene in a region below $30$ meV.

The screening effects play a very essential role in the gap generation in few-layer graphene.
They weaken the electron-electron interactions and increase very sharply with the number of layers $n$
in ABC-stacked multilayer graphene. The physical reason for this is connected with the density of states
$D(\varepsilon)\sim \varepsilon^{(2-n)/n}$ at energy $\varepsilon$ in gapless rhombohedral graphene,
which implies that, as $\varepsilon \to 0$, the density of states vanishes for single layer graphene,
is constant for bilayer graphene, and diverges for three and higher $n$-layer graphene. Since the
behavior of the static polarization function $\Pi(k)$ for $k \to 0$ is connected with the density of states at $\varepsilon=0$, it is natural to expect that the static polarization function
should diverge for $k \to 0$ in three and higher $n$-layer rhombohedral graphene. Indeed, a recent calculation \cite{Min-polarization} found that in undoped ABC-stacked multilayer graphene the static polarization function $\Pi(k) \sim k^{2-n}$ and, consequently, diverges for $n\ge 3$ if $k \to 0$. The static polarization function in rhombohedral multilayer graphene was studied also in Ref. \cite{Gelderen}.

Therefore, although the band flattening makes favorable the gap generation in multilayers with large $n$,
stronger screening effects weaken the electron-electron interaction. Consequently, {\it a priori} it is not clear whether the experimentally observed growth of the gap in undoped chiral multilayer graphene
will continue for $n \ge 4$. Clearly, to determine how the gap evolves with the number of layers in rhombohedral graphene is of great fundamental and practical interest. This problem provides the main motivation for the present paper. We would like to mention also that the exact diagonalization method
was employed in Ref. \cite{Hatsugai} to study the gap generation in ABC-stacked multilayer graphene in a magnetic field and it was found there that the gap decreases monotonously with $n$ starting from $n=2$, but what happens in the absence of a magnetic field is an open problem.

The paper is organized as follows. We set up the model in Sec.\ref{2}. The dynamical polarization function is numerically calculated and its fitting function is found in Sec.\ref{polarization}. We derive the gap equation and obtain its solution in Sec.\ref{gjgap}. The main results are
summarized and discussed in the conclusion.

\section{Model}
\label{2}

\begin{figure}[htp!]
\includegraphics[scale=0.2]{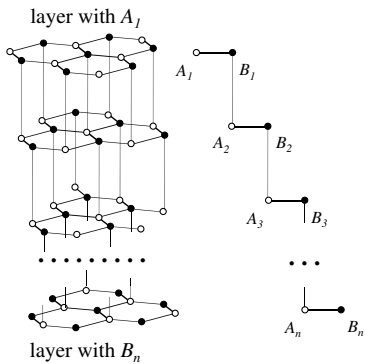}
\caption{Lattice structure of ABC-stacked multilayer graphene \label{fig-schematic}}
\end{figure}

Neglecting the trigonal warping effects, the low energy electron Hamiltonian in chiral multilayer graphene with $n \ge 2$ layers is given
by \cite{Min,Barlas,Cote,Nakamura}
\be
H_0=\sum_{\xi,s}\int d^2\mathbf{x}\,\Psi^+_{\xi s}(\mathbf{x})\left[\,-\xi^na_n\left(
\begin{array}{cc} 0 & (\hat{k}_-)^n\\
(\hat{k}_+)^n & 0 \end{array}
\right)
\right]\Psi_{\xi s}(\mathbf{x})\,,
\label{free-Hamiltonian}
\ee
where $\hat{k}_{\pm}=\hat{k}_x\pm i\hat{k}_y$, $\hat{\bf k}$ is the canonical momentum operator
(we set the Planck constant $\hbar=1$), $a_n=\gamma_1(v_F/\gamma_1)^n$, $v_F \sim 10^6 m/s$ is
the Fermi velocity in graphene, and $\gamma_1 \approx 0.39$ eV. The low energy effective Hamiltonian (\ref{free-Hamiltonian}) can be utilized for momenta $k$ up to $k_{W}=\gamma_1/v_F$. The two-component spinor field  $\Psi_{\xi s}$ carries the valley ($\xi=\pm$ for the $K$ and $K^{\prime}$ valleys,
respectively) and spin ($s=\pm$) indices. For the ABC-stacked multilayer graphene, the
low-energy electron states are located only on the outermost layers (see Fig.\ref{fig-schematic}),
which we will denote as layers $1$ and $n$ in what follows. Further, we use the standard convention
for wave functions:
$\Psi_{+s}^T=(\psi_{+A{_1}}, \psi_{+B{_n}})_s$, whereas, $\Psi_{-s}^T = (\psi_{-B{_n}},
\psi_{-A{_1}})_s$. Here $A_1$ and $B_n$ correspond to those sublattices in the outermost
layers $1$ and $n$, respectively, which are relevant for the low-energy dynamics. The effective
Hamiltonian (\ref{free-Hamiltonian}) is valid up to energies $\gamma_1/4\approx
0.1~\mbox{eV}$. Note that for $n=2$  Hamiltonian (\ref{free-Hamiltonian})
coincides  exactly with the low-energy Hamiltonian of bilayer graphene.

Using the free Hamiltonian (\ref{free-Hamiltonian}), we easily find the free quasiparticle
propagator in momentum space  at fixed valley $\xi$ and spin $s$
\be
S_{\xi s}(\bk,\omega)=\frac{1}{\omega^2-a^2_nk^{2n}}\left[\begin{array}{cc} \omega &
-\xi^n a_n(k_-)^n\\-\xi^n a_n(k_+)^n  & \omega\end{array}\right].
\label{free-propagator}
\ee
 Then taking into account the dynamically generated gap $\Delta_{\xi s}$, we have the following full quasiparticle propagator at fixed valley $\xi$ and spin $s$:
\be
G_{\xi s}(\bk,\omega)=\frac{1}{\omega^2-a^2_nk^{2n}-\Delta^2_{\xi s}}\left[
\begin{array}{cc} \omega+\Delta_{\xi s} & -\xi^n a_n(k_-)^n\\
-\xi^na_n(k_+)^n  & \omega-\Delta_{\xi s} \end{array}\right].
\label{full-propagator}
\ee
The Coulomb interaction between the electrons is described by the following interaction Hamiltonian:
\ba
H_{\rm int}&=&\frac{e^2}{2\kappa}\int\hspace{-1.0mm}
d^3\mathbf{r}d^3\mathbf{r}^{\prime}\frac{n_{\rm el}(\mathbf{r})n_{\rm el}(\mathbf{r}^{\prime})}{|\mathbf{r}-
\mathbf{r}^{\prime}|},
\label{Coulomb-interaction}
\ea
where $\kappa$ is the dielectric constant (we use $\kappa=4$ in the present paper),
$n_{\rm el}(\mathbf{r})=\delta(z-(n-1)d/2)\rho_1(\mathbf{x})+\delta(z+(n-1)d/2)\rho_n(\mathbf{x})$
is the three dimensional electron density in the outermost layers of ABC-stacked $n$-layer
graphene ($d \simeq 0.35$nm is the distance between the two neighbor layers), and
two-dimensional charge densities $\rho_1(\mathbf{x})$ and $\rho_n(\mathbf{x})$ in the outermost layers 1
and $n$ are
\be
\rho_1(\mathbf{x})=\Psi^+(\mathbf{x})P_1\Psi(\mathbf{x})\,,\quad \rho_n(\mathbf{x})=\Psi^+(\mathbf{x})P_n\Psi(\mathbf{x})\,,
\label{density}
\ee
where $P_1=(1+\xi\tau^3)/2$ and $P_n=(1-\xi\tau^3)/2$ are projectors on states in the layers
$1$ and $n$, respectively, and the Pauli matrix $\tau^3$ acts on layer components.

Integrating over $z$ and $z^{\prime}$ in Eq. (\ref{Coulomb-interaction}), one can rewrite the
interaction Hamiltonian as follows:
\ba
H_{\rm int}=\frac{1}{2}\int \hspace{-1.0mm}d^2\mathbf{x}d^2\mathbf{x}^{\prime}\left[V(\mathbf{x}-\mathbf{x}^{\prime})
\left(\rho_1(\mathbf{x})\rho_1(\mathbf{x}^{\prime})+\rho_n(\mathbf{x})\rho_n(\mathbf{x}^{\prime})\right)\hspace{-1.0mm}
+2V_{1n}(\mathbf{x}-\mathbf{x}^{\prime})\rho_1(\mathbf{x})\rho_n(\mathbf{x}^{\prime})\right].
\label{Coulomb-interaction-1}
\ea
Here the potential $V(\mathbf{x})$ describes the intralayer interactions and, therefore, coincides with the bare potential in monolayer graphene
whose Fourier transform is given by $\tilde{V}(k)={2\pi e^2}/{\kappa k}$, $k=|\mathbf{k}|$. The potential $V_{1n}$ describes the interlayer
electron interactions. Its Fourier transform is $\tilde{V}_{1n}(k)=({2\pi e^2}/{\kappa})({e^{-k(n-1)d}}/{k})$. Note that the form of the
interaction Hamiltonian (\ref{Coulomb-interaction-1}) coincides with that in bilayer graphene \cite{GGM}.

\section{Dynamical polarization function}
\label{polarization}

Taking into account gap $\Delta \ne 0$ qualitatively changes for $n \ge 3$ the behavior of
the polarization function for $k < \Delta$ because it no longer diverges for $k \to 0$ in
this case. Consequently, it is very important to calculate the polarization function for gapped
quasiparticles in order to get a correct gap equation. Further, the investigation of the gap
generation in bilayer graphene \cite{GGJM,GGMSh-mixing} showed that the dynamical screening
leads to three times larger gap compared to the case of the static screening. Therefore, we
will calculate the dynamical polarization function for gapped quasiparticles in rhombohedral
multilayer graphene and use it in the gap equation.

The dynamical polarization function $\Pi_{jk}$ describes electron density correlations on
the layers $j,k=1,n$,
\be
\delta(\omega+\omega^{\prime})\delta(\mathbf{k}+\mathbf{k}^{\prime})\Pi_{jk}(\omega,\mathbf{k})=
-i\langle 0|\rho_j(\omega,\mathbf{k})\rho_k(\omega^{\prime},\mathbf{k}^{\prime})|0\rangle \,.
\label{polarization-functions}
\ee
There are two independent functions, $\Pi_{11}=\Pi_{nn}$ and $\Pi_{1n}=\Pi_{n1}$. Taking into account the screening effects, the bare electron-electron interactions transform into
\ba
\hat{V}_{\mbox{\scriptsize eff}}=\hat{V}\cdot\frac{1}{1+\hat{V}\cdot\hat{\Pi}}=
\left(\begin{array}{cc}\tilde{V}_{\mbox{\scriptsize eff}}(k)&\tilde{V}_{1n\,\mbox{\scriptsize eff}}(k)
\\ \tilde{V}_{1n\,\mbox{\scriptsize eff}}(k)& \tilde{V}_{\mbox{\scriptsize eff}}(k)\end{array}\right),
\quad \hat{V}=\left(\begin{array}{cc}\tilde{V}(k)&\tilde{V}_{1n}(k)
\\ \tilde{V}_{1n}(k)& \tilde{V}(k)\end{array}\right),\quad \hat{\Pi}=\left(\begin{array}{cc}
\Pi_{11}(k)&\Pi_{1n}(k)
\\ \Pi_{1n}(k)& \Pi_{11}(k)\end{array}\right),
\ea
with
\ba
&&\tilde{V}_{\mbox{\scriptsize eff}}(\omega,k)=\frac{2\pi e^2}{\kappa}\,\frac{k+\frac{2\pi
e^2}{\kappa}\Pi_{11}(1-e^{-2(n-1)kd})} { \left[k+\frac{2\pi e^2}{\kappa}(\Pi_{11}+\Pi_{1n})
(1+e^{-(n-1)kd})\right]\left[k+\frac{2\pi e^2}{\kappa}(\Pi_{11}-\Pi_{1n})
(1-e^{-(n-1)kd})\right]}\,, \label{interaction-effective}\\
&& \tilde{V}_{1n\,\mbox{\scriptsize eff}}(\omega,k)=\frac{2\pi e^2}{\kappa}\,\frac{ke^{-(n-1)kd}
-\frac{2\pi e^2}{\kappa}\Pi_{1n}(1-e^{-2(n-1)kd})} {\left[k+\frac{2\pi e^2}{\kappa}
(\Pi_{11}+\Pi_{1n})(1+e^{-(n-1)kd})\right]\left[k+\frac{2\pi e^2}{\kappa}(\Pi_{11}-\Pi_{1n})
(1-e^{-(n-1)kd})\right]}\,.
\label{interaction-ND}
\ea
Since $\Pi_{11}$ and $\Pi_{1n}$ depend on $\omega$, the effective
interactions $\tilde{V}_{\mbox{\scriptsize eff}}$ and $\tilde{V}_{1n\,\mbox{\scriptsize eff}}$
depend on it too.

Using the ultraviolet cut-off $k_{W}=\gamma_1/v_F$ for our low-energy model, we find $k_{W}d=0.2$.
Therefore, we can approximate $e^{-(n-1)kd} \approx 1$ for $n \le 6$. Then we find
\ba
&&\tilde{V}_{\mbox{\scriptsize eff}}(\omega,k)=\frac{2\pi e^2}{\kappa}\,\frac{k+\frac{2\pi
e^2}{\kappa}\Pi_{11}(1-e^{-2(n-1)kd})} {k\left[k+\frac{4\pi e^2}{\kappa}\Pi\right]}\,,
\label{interaction-D}\\
&& \tilde{V}_{1n\,\mbox{\scriptsize eff}}(\omega,k)=\frac{2\pi e^2}{\kappa}\,\frac{k-\frac{2\pi
e^2}{\kappa}\Pi_{1n}(1-e^{-2(n-1)kd})} {k\left[k+\frac{4\pi e^2}{\kappa}\Pi\right]},
\label{interaction-approximate}
\ea
where
\be
\Pi(\omega,\mathbf{k})\equiv \Pi_{11}(\omega,\mathbf{k})+\Pi_{1n}(\omega,\mathbf{k})\label{pfdefed}
\ee
is the dynamical polarization function. Using Eq. (\ref{polarization-functions}), we find that
the functions $\Pi_{jk}$ are given by
\be
\Pi_{jk}(\omega,\mathbf{p})=i\int \frac{d\omega'
d^2k}{(2\pi)^3}\,\mbox{tr}\,\left[\,P_{j}\,G(\omega',\mathbf{k}-\mathbf{p}/2)\,P_{k}\,
G(\omega+\omega',\mathbf{k}+\mathbf{p}/2)\,\right],\quad j,k=1,n,
\label{p11}
\ee
where trace includes the summation over valley and spin indices. Clearly, the dynamical
polarization function equals
\begin{equation}
\Pi(\omega,\bp)=i\int \frac{d\omega' d^2k}{(2\pi)^3}\,\tr \left[ P_1 G(\omega'-\omega/2, \bk-\bp/2)G(\omega'+\omega/2,\bk+\bp/2)\right].
\label{pfdef2}
\end{equation}
Since the aim of the present paper is to study how the gap generation depends on the
number of layers $n$, for the sake of simplicity, we will assume in our calculation of
the polarization function that the gap $\Delta_{\xi s}$ does not depend on $\xi$ and $s$.
Using Eq. (\ref{p11}), the full electron propagator (\ref{full-propagator})
and integrating over $\omega'$, we obtain
\begin{equation}
\Pi_{1n}(\omega,\mathbf{p})=\Pi_{n1}(\omega,\mathbf{p}) = a_n^2\int\frac{d^2k}{(2\pi)^2} \frac{(\epsilon_-+\epsilon_+)\left[(\bk-\frac{\bp}{2})^n_+ (\bk+\frac{\bp}{2})^n_- +
(\bk-\frac{\bp}{2})^n_-(\bk+\frac{\bp}{2})^n_+ \right]}
{\epsilon_-\epsilon_+\left(\omega^2-(\epsilon_-+\epsilon_+)^2\right)},
\label{pfp1}
\end{equation}
\begin{equation}
\Pi_{11}(\omega,\mathbf{p})=\Pi_{nn}(\omega,\mathbf{p})=-2\int\frac{d^2k}{(2\pi)^2}
\frac{(\epsilon_-+\epsilon_+)\left(\epsilon_-\epsilon_+-\Delta^2 \right)}
{\epsilon_-\epsilon_+\left(\omega^2-(\epsilon_-+\epsilon_+)^2\right)},\quad \epsilon_\pm
=\sqrt{\Delta^2+\left(a_n|\bk\pm\frac{\bp}{2}|^n\right)^2}.
\label{pfp2}
\end{equation}
Unfortunately, in the gapped case, the integrals in Eqs. (\ref{pfp1}) and (\ref{pfp2}) cannot
be calculated analytically. For $\Delta=0$, $\epsilon_{\pm}$ simplifies to
\be
\epsilon^{\Delta=0}_{\pm}=a_n|\bk\pm\frac{\bp}{2}|^n=a_n(k^2\pm kp\cos\theta+p^2/4)^{n/2},
\ee
where $\theta$ is the angle between $\bk$ and $\bp$. However, even in this case the integrals in Eqs. (\ref{pfp1}) and (\ref{pfp2}) can be
analytically calculated only for small even integers
$n=2,4,6$. For example, we checked that for $n=2$ Eqs. (\ref{pfp1}),(\ref{pfp2}) are
in agreement with corresponding expressions for bilayer graphene.

\subsection{Polarization function $\Pi$}
\label{pfnzg}

\begin{figure}[htp!]
\includegraphics[scale=0.40]{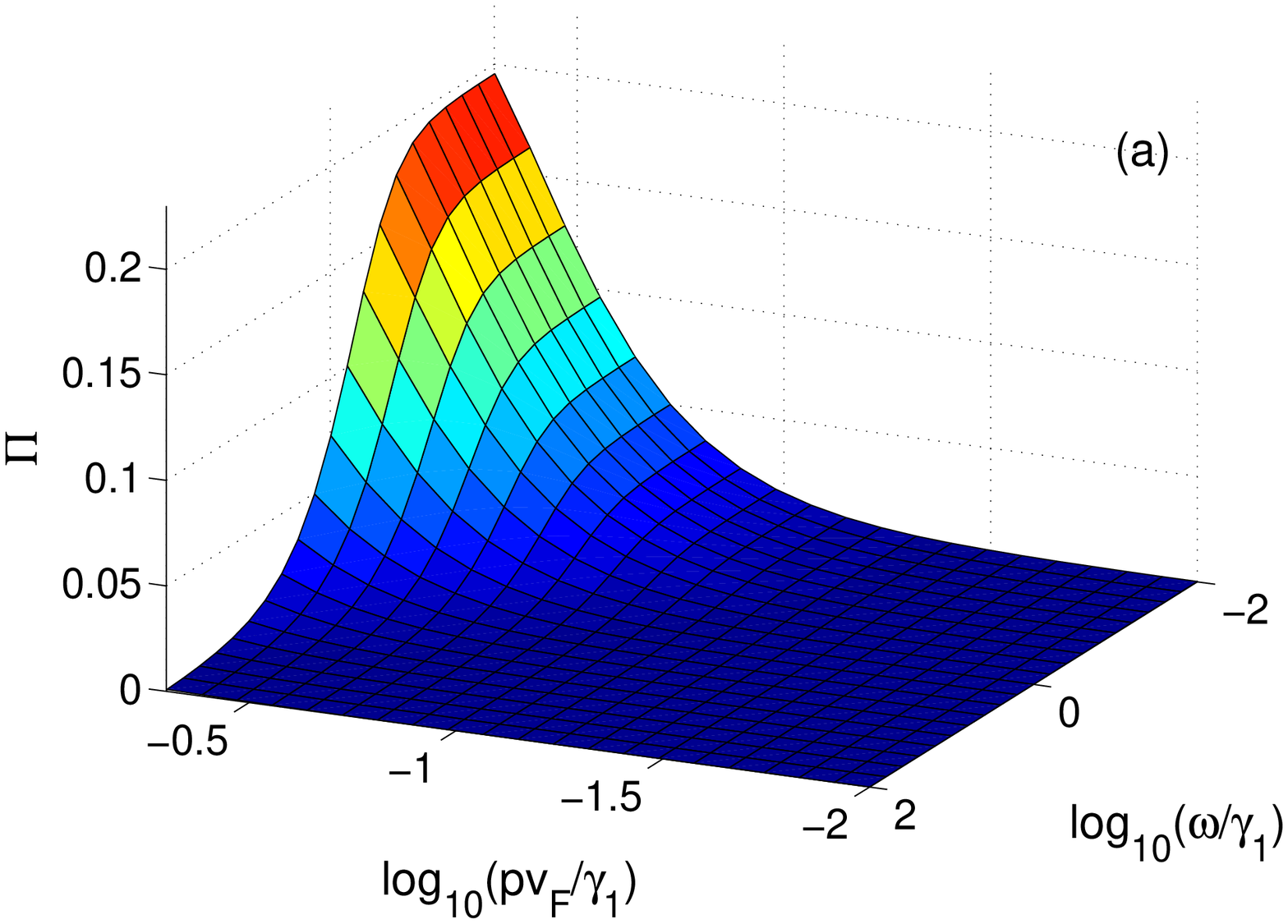}
\includegraphics[scale=0.40]{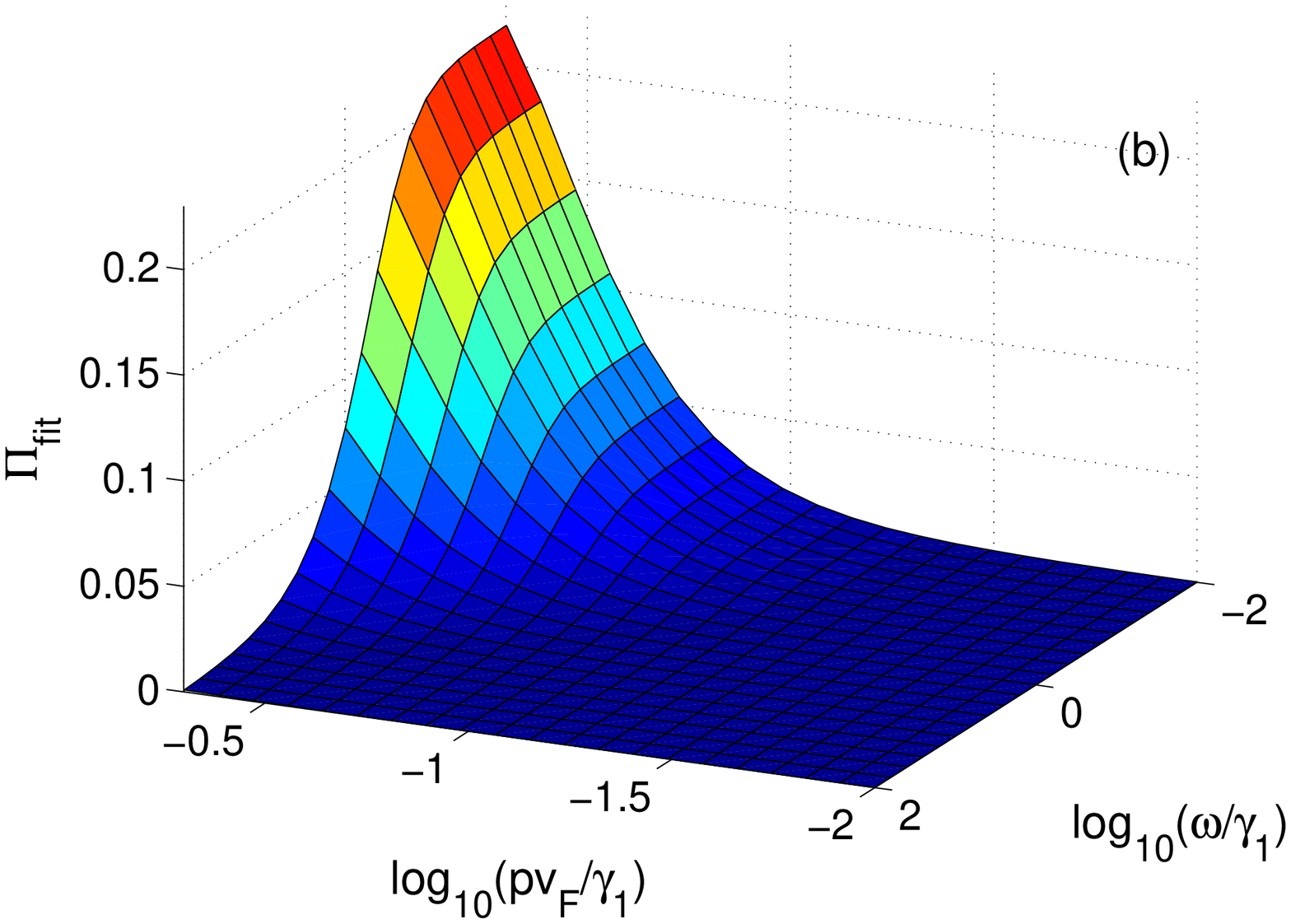}
\includegraphics[scale=0.40]{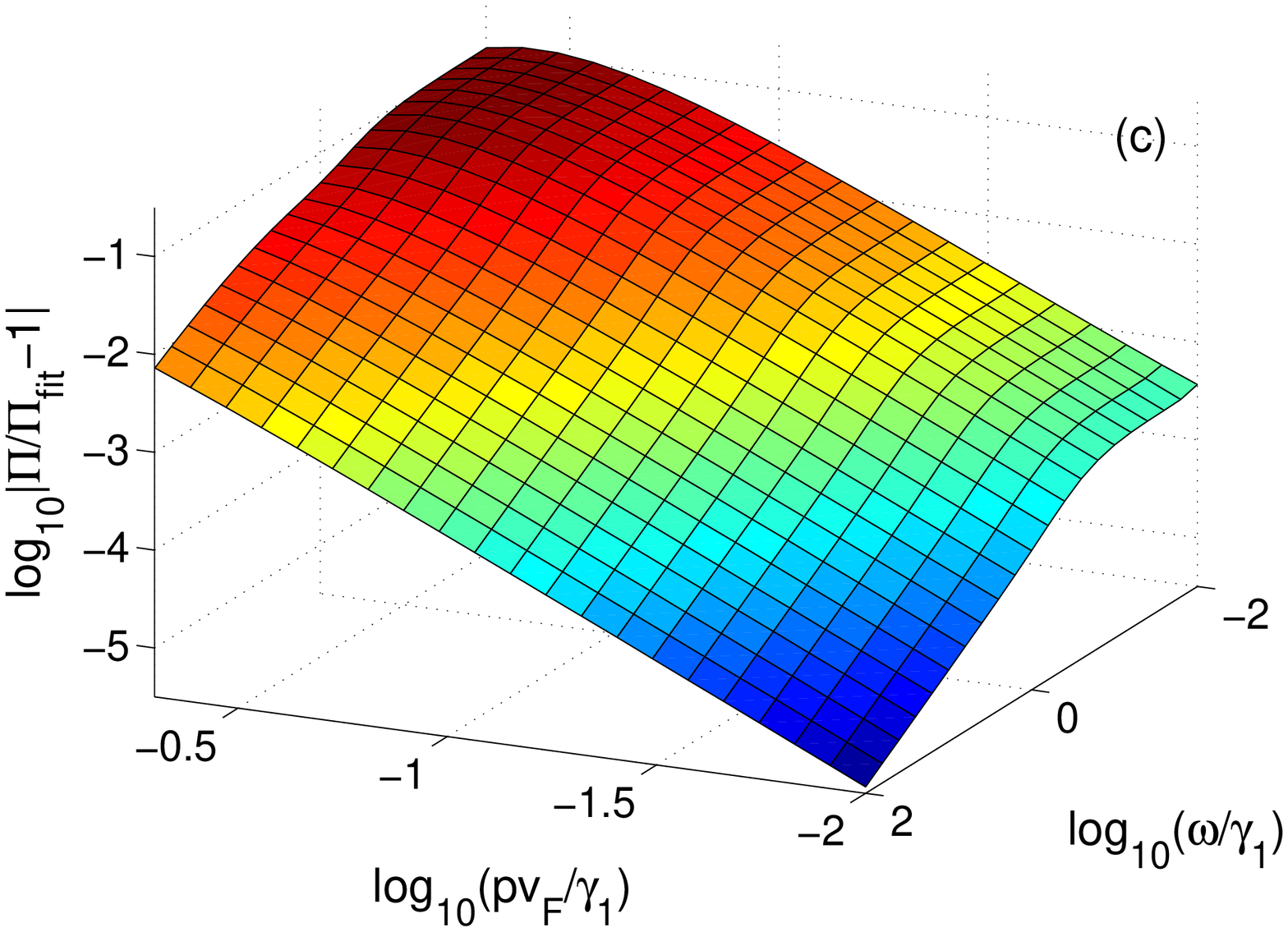}
\caption{(Color online) (a) The numerically calculated dynamical polarization function \refer{pfdefed}.
(b) The fitting function \refer{pffitimp1}. (c) The logarithm of the modulus of the relative difference
of polarization functions \refer{pfdefed} and \refer{pffitimp1}.
In all plots, we take $n=3$ and set $\Delta/\gamma_1=0.1$. \label{fig2}}
\end{figure}

In the gap equation, we finally need the polarization function in the Euclidean region $\omega^2<0$
(this corresponds to the Wick rotation). Using Eqs. (\ref{pfp1}) and (\ref{pfp2}), we have numerically computed the dynamical polarization function (\ref{pfdefed}) in the Euclidean region and plot it for
$n=3$ and $\Delta/\gamma_1=0.1$ in Fig.\ref{fig2}(a) as a function of $\mbox{log}_{10}(pv_F/\gamma_1)$
and $\mbox{log}_{10}(\omega/\gamma_1)$.
In order to solve the gap equation, it is essential to have an analytic fit to the numerically
computed dynamical polarization function. To find such a fit, we will determine first the small and large momentum asymptotic of the dynamical
polarization function and them match them using an interpolating fitting function. We begin with
the small momentum expansion. Expanding $\Pi(\omega,p)$ up to $p^4$, the integral in Eq. (\ref{pfdef2})
can be analytically calculated. We find
\be
\Pi(\omega,p\to0,\Delta)= \frac{p^2n}{2\pi}\left\{\frac{\Delta}{\omega^2}
-\frac{(4\Delta^2-\omega^2)\left[\pi-2\arcsin\left(\frac{2\Delta}{\sqrt{\omega^2+4\Delta^2}}
\right)\right]}{4|\omega|^3}\right\} + {\cal O}(p^4)\equiv \frac{p^2}{b_n}+ {\cal O}(p^4).
\label{pfspappres2}
\ee
For large momentum asymptotic of $\Pi$, using Eqs. (\ref{pfdef2}), (\ref{pfp1}), and (\ref{pfp2}),
we obtain
\be
\Pi(\omega,p\to\infty,\Delta)=\frac{c_n}{a_n p^{n-2}}+{\cal O}\left(\frac{1}{p^n}\right),\quad\quad c_2=\frac{\ln2}{\pi},\quad c_3=0.24,\quad
c_4=\frac{1}{4},\quad
c_5=0.29,\quad c_6=\frac{2\sqrt{3}}{9}.
\label{polarization-function-asymptotic}
\ee
Combining Eqs. (\ref{pfspappres2}) and (\ref{polarization-function-asymptotic}), it is not difficult
to find a simple yet accurate fitting function that interpolates between the small and large
momentum expansions
\be
\Pi_{\mbox{\scriptsize fit}}(\omega,p,\Delta)=\frac{p^2}{b_n+a_np^n/c_n}.
\label{pffitimp1}
\ee
Comparing this function with the numerically computed dynamical polarization function, we find
that function (\ref{pffitimp1}) agrees very well with the exact polarization function. This
fitting function for $n=3$ is plotted in Fig.\ref{fig2}(b) for
$\Delta/\gamma_1=0.1$ as a function of $\log_{10}(pv_F/\gamma_1)$ and $\log_{10}(\omega/\gamma_1)$.
In Fig.\ref{fig2}(c), we plot the logarithm of the modulus of the relative difference
of the numerically computed $\Pi$ and the fitting function (\ref{pffitimp1}). One can see that the
deviation of the fitting function from the exact function is well below 1\% except only in very small frequency and momentum region, where deviation can reach up to 10\% but not more than that. Since this region is very small compared to the whole integration region, we believe that the gaps calculated using
the fitting function (\ref{pffitimp1}) in the next section are accurate enough.

\subsection{Polarization function $\Pi_{11}$}

\begin{figure}[htp!]
\includegraphics[scale=0.40]{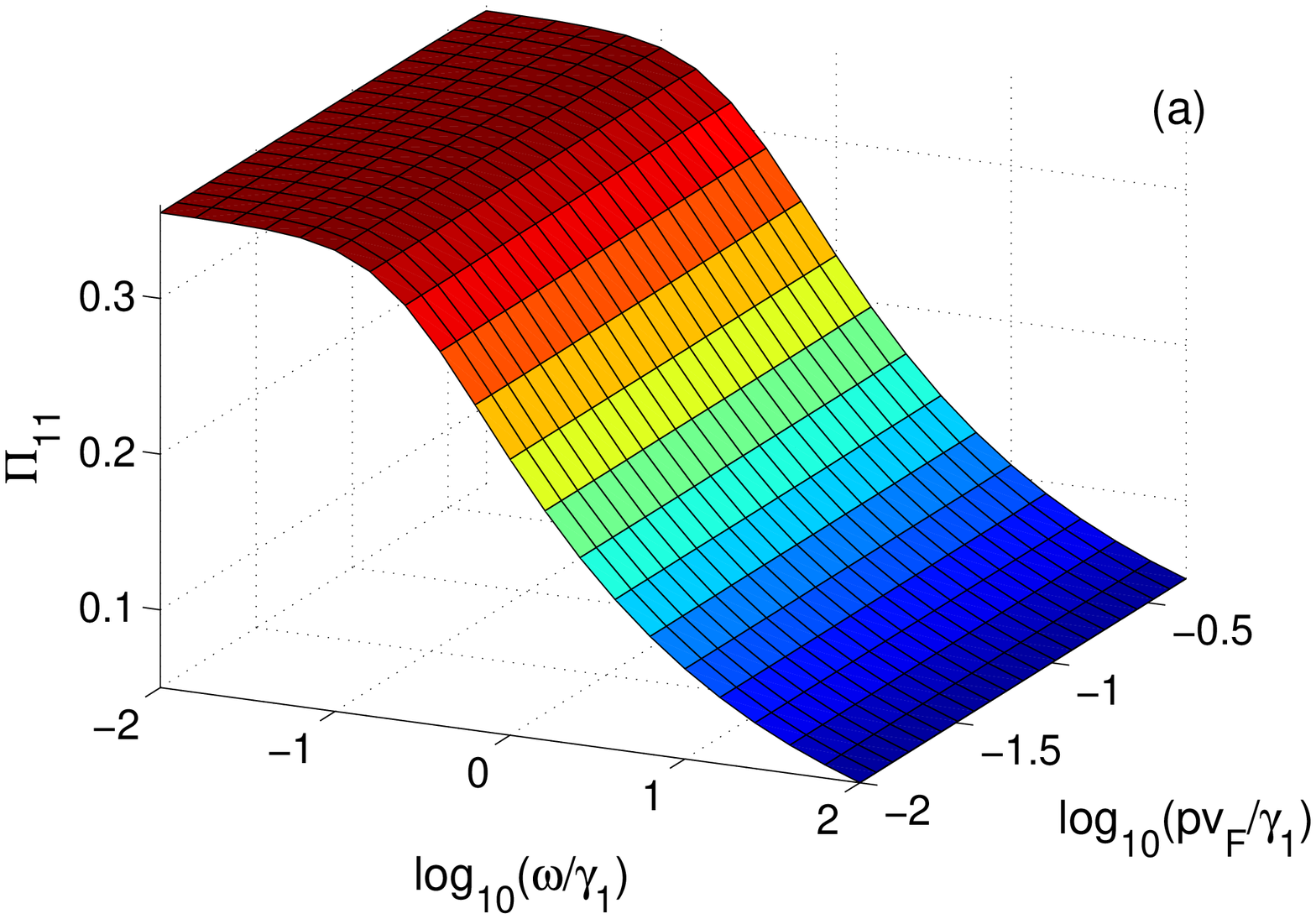}
\includegraphics[scale=0.40]{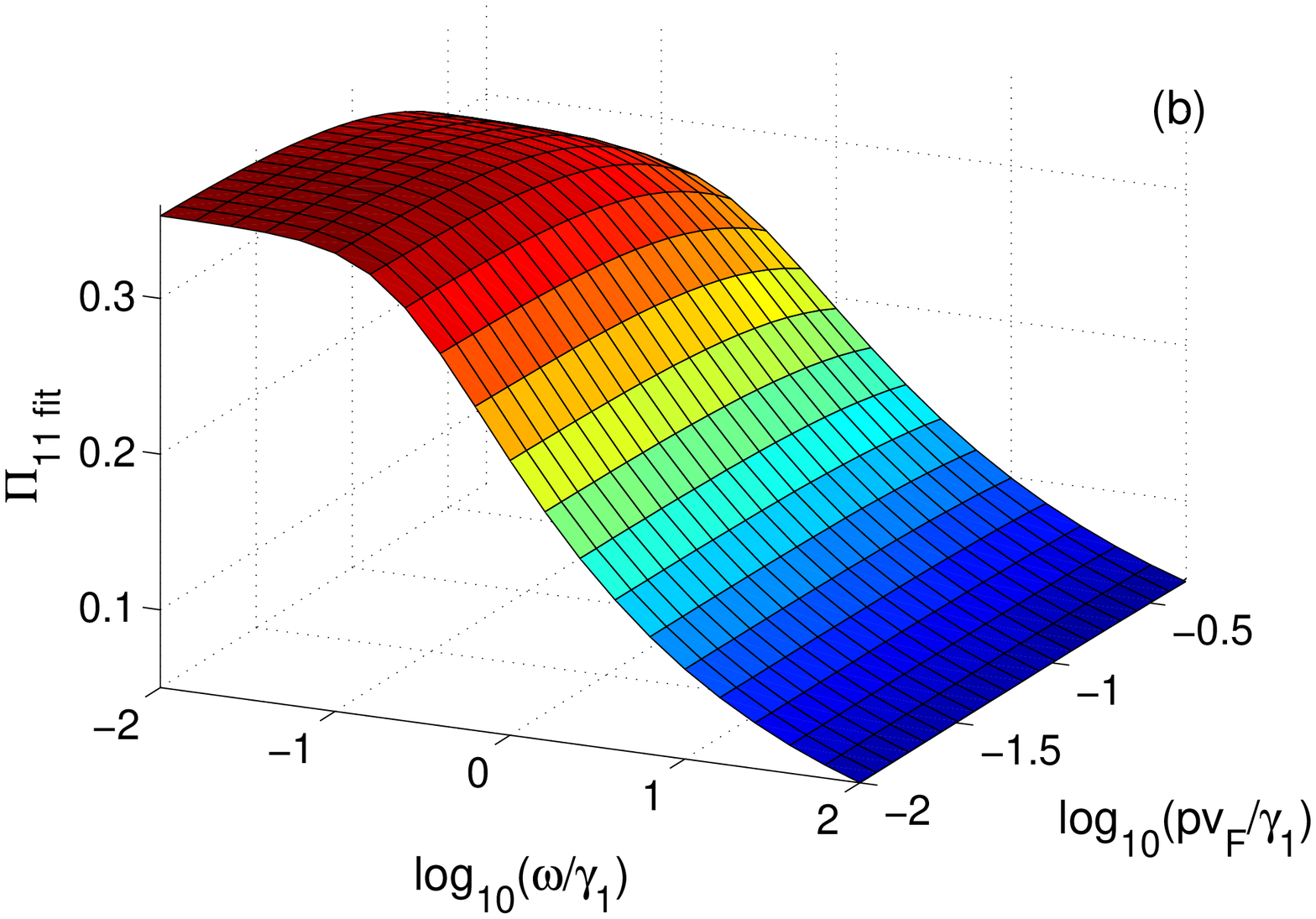}
\includegraphics[scale=0.40]{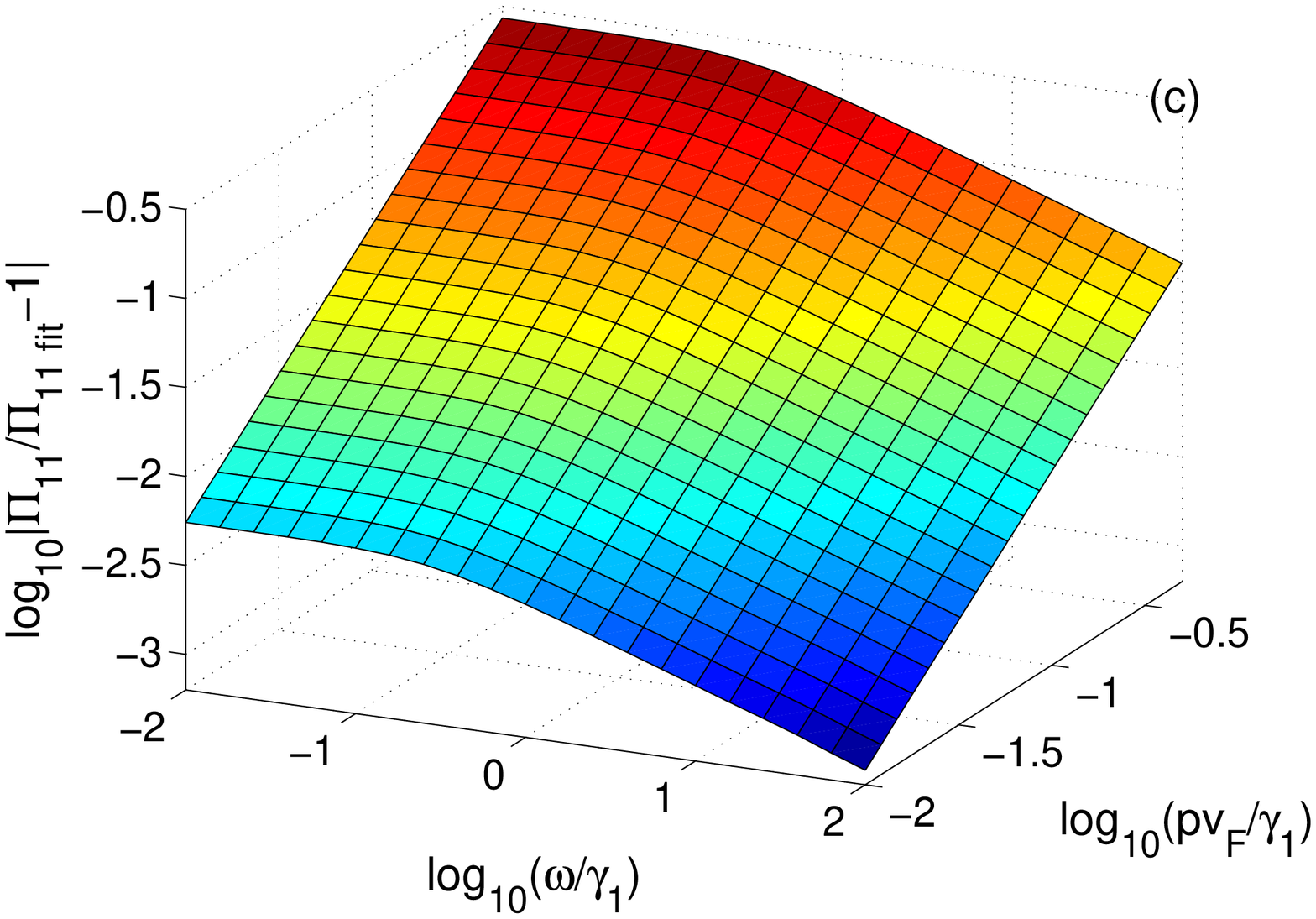}
\caption{(Color online) (a) The exact polarization function $\Pi_{11}$ defined in Eq. \refer{pfp2}.
(b) The fitting function \refer{pf11fitimp1}. (c) The logarithm of the modulus of the relative difference
of polarization functions \refer{pfp2} and \refer{pf11fitimp1}. In all plots, we take $n=3$ and set $\Delta/\gamma_1=0.1$.
\label{pf11fig}}
\end{figure}

The effective interactions (\ref{interaction-D}) and (\ref{interaction-approximate}) in addition to
$\Pi$ contain also functions $\Pi_{11}$ and $\Pi_{1n}$. We will see in the next section that we need
to know only $\Pi_{11}$ for $n \ge 3$ in order to determine the gap for the state which we consider.
We have numerically computed $\Pi_{11}$ and plot it in Fig.\ref{pf11fig}(a) for $\Delta/\gamma_1=0.1$
as a function of $\mbox{log}_{10}(v_Fp/\gamma_1)$ and $\mbox{log}_{10}(\omega/\gamma_1)$. In order
to find its fitting function, we follow the same procedure as in the case of $\Pi$. At first, using Eq. (\ref{pfp2}), we find the following small
momentum expansion of $\Pi_{11}$:
\be
\Pi_{11}(\omega,p\to0,\Delta)=\frac{\Delta^{2/n -1}}{2\pi^{3/2} na_n^{2/n}}\left(\frac{4\Delta^2+\omega^2}
{4\Delta^2}\right)^{1/n}\Gamma\left(1+\frac{1}{n}\right)
\Gamma\left(\frac{n-2}{2n}\right)F\left(\frac{1}{2},1+\frac{1}{n},\frac{3}{2};-\frac{\omega^2}{4\Delta^2}
\right)+ {\cal O}(p^4)\equiv \frac{1}{f_n}+ {\cal O}(p^4),
\label{pi11ser}
\ee
where $F(a,b,c;z)$ is the Gauss hypergeometric function. The large momentum asymptotic of $\Pi_{11}$ coincides with that of $\Pi$ and is also given by Eq. (\ref{polarization-function-asymptotic})
\be
\Pi_{11}(\omega,p\to\infty,\Delta)=\Pi(\omega,p\to\infty,\Delta)= \frac{c_n}{a_n p^{n-2}}+{\cal O}\left(\frac{1}{p^n}\right).
\label{pi11asy}
\ee
Then we find the following fitting function that matches both asymptotics (\ref{pi11ser}) and (\ref{pi11asy}):
\be
\Pi_{11\,\mbox{\scriptsize fit}}(\omega,p,\Delta)=\frac{1}{f_n+a_np^{n-2}/c_n}.
\label{pf11fitimp1}
\ee

In  Fig. \ref{pf11fig}(b), the fitting function (\ref{pf11fitimp1}) is plotted for $\Delta/\gamma_1=0.1$ as a function of
$\mbox{log}_{10}(pv_F/\gamma_1)$ and $\mbox{log}_{10}(\omega/\gamma_1)$.
In Fig. \ref{pf11fig}(c), we plot the logarithm of the modulus of the relative
difference of the numerically calculated $\Pi_{11}$ and the fitting function (\ref{pf11fitimp1}).
One can see that the deviation of the fitting function from the exact function is well below 10\% except in very small frequency and momentum region,
where deviation can be larger. However, similar to the case of $\Pi$, this deviation is expected to cause little effect in the computation of gap.

\section{The gap equation and its solution}
\label{gjgap}

The derivation of the gap equation for ABC-stacked $n$-layer graphene is completely analogous to
that for bilayer graphene $n=2$ considered in \cite{GGJM,GGMSh-mixing}. Our starting point is the Schwinger--Dyson equation for the full
quasiparticle propagator which in the Hartree-Fock approximation reads
\begin{eqnarray}
G^{-1}(t-t^\prime;\bm{r},\bm{r}^\prime)&=&S^{-1}(t-t^\prime;\bm{r},\bm{r}^\prime)
- iG(t-t^\prime;\bm{r},\bm{r}^\prime) V_{\rm eff} (t-t^\prime;\bm{r}-\bm{r}^\prime)\nonumber\\
&-& i\left[P_{1}G(t-t^\prime;\bm{r},\bm{r}^\prime)P_{n}
+P_{n}G(t-t^\prime;\bm{r},\bm{r}^\prime)P_{1}\right]
{V}_{\rm IL}(t-t^\prime;\bm{r-\bm{r}^\prime})\nonumber\\
&-&\frac{i}{2}(\,P_{1}-P_n\,)\,\mbox{tr}\left[\,(\,P_{1}-P_{n}\,)G(0;0)\,\right]
\tilde{V}_{\rm IL}^{\rm bare}(0)  \delta(t-t^\prime)\delta(\bm{r}-\bm{r}^\prime),
\label{SD-equation}
\end{eqnarray}
where $V_{\rm IL}(t-t^{\prime},\mathbf{r}-\mathbf{r}^{\prime})=V_{\rm 1n\,eff}(t-t^{\prime},
\mathbf{r}-\mathbf{r}^{\prime})-V_{\rm eff}(t-t^{\prime},
\mathbf{r}-\mathbf{r}^{\prime})$ and $\tilde{V}^{\rm bare}_{IL}(0)=-2\pi e^2d/\kappa$.

The experimental results in Refs.\cite{Bao,Lee} suggest that the gapped state in ABC-stacked
trilayer graphene in the absence of external electromagnetic fields is well described (like
in bilayer graphene) by the layer antiferromagnetic (LAF) solution $\Delta_{\xi s}=\xi s \Delta$ (the
experimental data in \cite{Kurganova} indicate too on a spin unpolarized ground state of
trilayer graphene). We assume that this is true for $n \ge 4$ as well. For the LAF solution,
the last term in Eq. (\ref{SD-equation}) vanishes due to trace over spin indices. Then, using Eqs. (\ref{free-propagator}) and (\ref{full-propagator}), we find that the Schwinger--Dyson
equation (\ref{SD-equation}) leads to the following equation in momentum space for the LAF
gap $\Delta$ (it is easy to check also that the third term on the right-hand side of
Eq. (\ref{SD-equation}) does not contribute to the gap equation):
\be
\Delta(\Omega,p)=\int \frac{d\omega d^2k}{(2\pi)^3}
\frac{\Delta(\omega,k)}{\omega^2+\Delta^2+(a_n k^n)^2}\,\tilde{V}_{\mbox{\scriptsize eff}}(\Omega-\omega,p-k),
\label{gapeq}
\ee
where $\tilde{V}_{\mbox{\scriptsize eff}}$ is given by Eq. (\ref{interaction-D}).
For the purpose of this paper it suffices to solve Eq. (\ref{gapeq}) in the simplest constant gap approximation. This means that we  set the external frequency and momentum to zero,
$\Omega=p=0$, in the gap equation and utilize the frequency and momentum independent gap.
However, we keep the dependence on $\omega,k$ in $\tilde{V}_{\mbox{\scriptsize eff}}$. The reason
is that since the static polarization function strongly overestimates
the screening effects at large $\omega$, it is essential to take into account the dynamical
polarization function in the analysis. We note that
the trigonal warping effects and the dependence of the gap on momentum neglected in
the constant gap approximation could significantly affect the gap magnitude. This important problem
will be considered elsewhere.

\begin{figure}[htp!]
\includegraphics[scale=0.4]{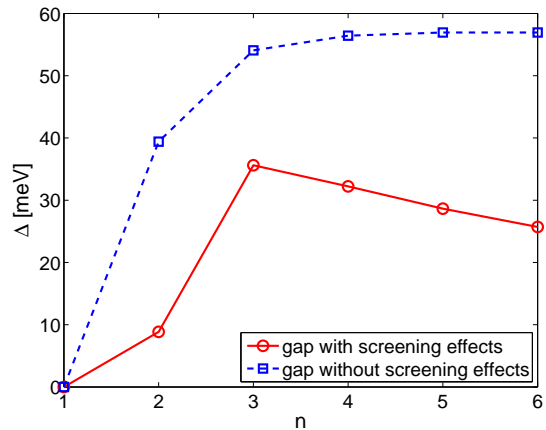}
\caption{(Color online) The gap as a function of $n$ with the screening
effects taken into account (red solid line) and without the screening effects (dashed blue line). \label{figgapj}}
\end{figure}

Using the fitting functions $\Pi_{\mbox{\scriptsize fit}}$ and
$\Pi_{11\,\mbox{\scriptsize fit}}$ given by Eqs. (\ref{pffitimp1}) and (\ref{pf11fitimp1}),
the gap equation (\ref{gapeq}) in the constant gap approximation takes the form
\be
1=\int \frac{d\omega d^2k}{(2\pi)^3}\frac{1}{\omega^2+\Delta^2+(a_nk^{n})^2}
\frac{2\pi e^2}{\kappa}\,\frac{k+\frac{2\pi
e^2}{\kappa}{\Pi}_{11\,\mbox{\scriptsize fit}}(1-e^{-2(n-1)kd})} {k\left[k+
\frac{4\pi e^2}{\kappa}{\Pi}_{\mbox{\scriptsize fit}}\right]}\,.
\label{gapeqscaled}
\ee
Since the distance between the two neighbor layers $d$ is very small, the factor $(1-e^{-2(n-1)kd})$
in Eq. (\ref{gapeqscaled}) is also very small when $n$ takes small values. Therefore for $n=2$,
we can set this factor to zero. This makes possible to consider the bilayer
case directly and allows us to avoid the complication that $\Pi_{11}$ for $n=2$ diverges in
the ultraviolet region in the effective low energy model and should be properly regularized.
We solve the gap equation (\ref{gapeqscaled}) numerically and plot the gap given by the
solid red line as a function of the number of layers $n$ in Fig. \ref{figgapj}.
The gap is measured in units of $\gamma_1$. Our main principal result is that the gap attains
maximum at $n=3$ and then decreases monotonically for $n \ge 4$. This suggests that ABC-stacked
trilayer graphene has the largest gap among chiral multilayer graphene systems.

It is instructive to solve also the gap equation (\ref{gapeqscaled}) neglecting the screening effects. Setting ${\Pi}_{\mbox{\scriptsize fit}}={\Pi}_{11\,\mbox{\scriptsize fit}}=0$, we solve Eq. (\ref{gapeqscaled}) and plot the corresponding solution given by the dashed blue line in Fig.\ref{figgapj}. As expected, the gaps are larger than those with the screening effects taken into account. Still
they are not much larger. The reason for this is connected with the fact that both $\Pi$
and $\Pi_{11}$ are positive, therefore, they somewhat compensate each other when we take them into account in the gap equation (\ref{gapeqscaled}). The most important difference between the two plots is that
the gap monotonously grows with $n$ when the screening effects are neglected. Consequently,
we conclude that the screening effects play a crucial role in the gap generation in ABC-stacked
multilayer graphene, compete with the band flattening, and single out trilayer graphene as chiral multilayer graphene with the largest gap.

\section{Conclusion}

We numerically calculated the dynamical polarization function in ABC-stacked $n$-layer graphene for
gapped quasiparticles and found its fitting function. Although the flattening of the low energy
electron bands suggests that the gap should increase with $n$, we found that the screening effects,
which sharply increase with the number of layers $n$ too, are more essential quantitatively. Solving
the self-consistent gap equation in the constant gap approximation taking into account the dynamical polarization function for gapped quasiparticles, we found that the gap attains its maximal value for trilayer graphene and then decreases monotonously for $n\ge 4$.

For ABC-stacked trilayer graphene, we obtained the gap $\Delta_3=36$ meV. Although we used the constant
gap approximation in order to solve the gap equation, our value is quite close to the most recent experimentally observed gap 25 meV \cite{Lee,LeRoy}. It is well known that taking into account the dependence of gap on momentum should decrease the gap. Certainly, it would be interesting to take into account this dependence in future studies of the gap generation in rhombohedral graphene. Furthermore,
since for tetra-layer graphene the gap $\Delta_4=32$ meV is only 10 \% less than the gap in trilayer graphene, it is important to check whether taking into account the trigonal warping effects and the dependence of the gap on momentum neglected in the present study will change our main conclusions. We would like to mention also that although the observed gap in the ABC-stacked trilayer graphene
$\Delta_3=25$ meV is almost of room temperature, the low-energy effective Hamiltonian (\ref{free-Hamiltonian}) is legitimate to use in the analysis because it is valid up to energy of
the order of $\gamma_1=0.39$ eV that is much larger than $\Delta_3$.

The results mentioned above are obtained for the dielectric constant $\kappa=4$. We have studied also
how gaps vary with $\kappa$ and found that while as expected their absolute values are pushed higher as $\kappa$ decreases, the general form of the dependence of $\Delta$ on $n$ is not changed and the relative ratios between gaps for different $n$ are not much changed. Varying $\kappa$, we found that the experimental value $\Delta_3=25$ meV is reproduced for $\kappa=5$. Interestingly, as a bonus result for this value of $\kappa$, we obtained $\Delta_2=3.4$ meV that is rather close to the experimental value and, as a benchmark for future experiments, we found that $\Delta_4=24.5$ meV and $\Delta_5=21.5$ meV. Although the dielectric constant $\kappa=5$ is unrealistic for experiments in suspended graphene, we expect that for smaller $\kappa$ taking into account  the dependence of the gaps on momentum will provide gaps with values comparable to the experimental ones.

Certainly, the most crucial check of our results would be an experimental determination of gaps in tetra- and higher $n$-layer rhombohedral graphene. Still we believe that our qualitative conclusion that the gap
in ABC-stacked multilayer graphene attains maximum at certain $n$ is consistent with the fact that naturally occurring rhombohedral graphite is a semimetal. If the gap were increased with $n$, then, obviously,
rhombohedral graphite would be an insulator.

\begin{acknowledgments}
We thank V.A.~Miransky and I.A.~Shovkovy for useful remarks.
The work of E.V.G and V.P.G. was supported partially by the European FP7 program Grant
No. SIMTECH 246937, the joint Ukrainian-Russian SFFR-RFBR Grant No.~F53.2/028, the grant
STCU \#5716-2 "Development of Graphene Technologies and Investigation of Graphene-based
Nanostructures for Nanoelectronics and Optoelectronics", and by the Program of Fundamental
Research of the Physics and Astronomy Division of the NAS of Ukraine. V.P.G. acknowledges
a collaborative grant from the Swedish Institute. J.J. is supported by the
Young Talent Starting Fund of Wuhan University \#202273006 and the SRFDP \#20130141120079. J.J. also thanks the hospitality of the
Department of Applied Mathematics, University of Western Ontario where during his stay part of the work was finished.
\end{acknowledgments}

\end{document}